\documentclass[pdftex,twocolumn,epjc3]{svjour3}  

\usepackage[T1]{fontenc}
\usepackage{graphicx}
\usepackage{mathptmx}      
\usepackage{flushend}
\usepackage[numbers,sort&compress]{natbib}
\usepackage[colorlinks,citecolor=blue,urlcolor=blue,linkcolor=blue]{hyperref}

\journalname{Eur. Phys. J. C}
\usepackage[utf8]{inputenc}
\usepackage{amsmath}
\usepackage{amssymb}

\makeatletter
\def\cl@chapter{\@elt {theorem}}
\makeatother
\usepackage[capitalise]{cleveref}
\usepackage{acronym}
\usepackage{xspace}

\acrodef{ML}{machine learning}
\acrodef{PF}{particle-flow}
\acrodef{GNN}{graph neural network}
\acrodef{LHC}{CERN Large Hadron Collider}
\acrodef{FCC}{Future Circular Collider}
\acrodef{MLPF}{machine-learned particle-flow}
\acrodef{GAN}{generative adversarial network}
\acrodef{GCN}{graph convolutional network}
\acrodef{PID}{particle identification}
\acrodef{kNN}{k-nearest neighbors}
\acrodef{LSH}{locality sensitive hashing}
\acrodef{GPU}{graphics processing unit}
\acrodef{FPGA}{field-programmable gate array}
\acrodef{ECAL}{electromagnetic calorimeter}
\acrodef{HCAL}{hadron calorimeter}
\acrodef{PU}{pileup}
\acrodef{IPU}{intelligence processing unit}

\newcommand{\ttbar}{\ensuremath{\mathrm{t}\overline{\mathrm{t}}}\xspace}

\newcommand{\PYTHIA} {{\textsc{pythia}}\xspace}
\newcommand{\DELPHES} {{\textsc{delphes}}\xspace}
\newcommand{\TENSORFLOW} {{\textsc{TensorFlow}}\xspace}
\newcommand{\pytorchgeometric} {{\textsc{Pytorch-Geometric}}\xspace}
\newcommand{\pt}{\ensuremath{p_{\mathrm{T}}}\xspace}
\newcommand{\GeV}{\ensuremath{\,\text{Ge\hspace{-.08em}V}}\xspace}
\newcommand{\TeV}{\ensuremath{\,\text{Te\hspace{-.08em}V}}\xspace}
\newcommand{\unit}[1]{\ensuremath{\text{\,#1}}\xspace}

\DeclareMathOperator*{\argmax}{arg\,max}
\begin{document}\sloppy
\title{MLPF: Efficient machine-learned particle-flow reconstruction using graph neural networks}

\author{Joosep Pata\thanksref{e1,addr1,addr2}
        \and
        Javier Duarte\thanksref{e2,addr3} 
        \and 
        Jean-Roch Vlimant\thanksref{e3,addr2}
        \and 
        Maurizio Pierini\thanksref{e4,addr4}
        \and
        Maria Spiropulu\thanksref{e5,addr2}
}

\thankstext{e1}{joosep.pata@cern.ch. Corresponding author. This work was partially carried out at Caltech.}
\thankstext{e2}{jduarte@ucsd.edu}
\thankstext{e3}{jvlimant@caltech.edu}
\thankstext{e4}{maurizio.pierini@cern.ch}
\thankstext{e5}{smaria@caltech.edu}
\institute{National Institute of Chemical Physics and Biophysics (NICPB), R\"{a}vala pst 10, 10143 Tallinn, Estonia\label{addr1}
          \and
          California Institute of Technology, Pasadena, CA 91125, USA\label{addr2}
          \and
          University of California San Diego, La Jolla, CA 92093, USA\label{addr3}
          \and
          European Center for Nuclear Research (CERN), CH 1211, Geneva 23, Switzerland\label{addr4}
}



\date{Received: 5 February 2021 / Accepted: 19 April 2021}

\maketitle 

\begin{abstract}
In general-purpose particle detectors, the \acl{PF} algorithm may be used to reconstruct a comprehensive particle-level view of the event by combining information from the calorimeters and the trackers, significantly improving the detector resolution for jets and the missing transverse momentum. 
In view of the planned high-luminosity upgrade of the \ac{LHC}, it is necessary to revisit existing reconstruction algorithms and ensure that both the physics and computational performance are sufficient in an environment with many simultaneous proton-proton interactions (\acl{PU}). 
Machine learning may offer a prospect for computationally efficient event reconstruction that is well-suited to heterogeneous computing platforms, while significantly improving the reconstruction quality over rule-based algorithms for granular detectors.
We introduce \acs{MLPF}, a novel, end-to-end trainable, machine-learned \acl{PF} algorithm based on parallelizable, computationally efficient, and scalable \aclp{GNN} optimized using a multi-task objective on simulated events. 
We report the physics and computational performance of the \acs{MLPF} algorithm on a Monte Carlo dataset of top quark-antiquark pairs produced in proton-proton collisions in conditions similar to those expected for the high-luminosity \ac{LHC}. 
The \acs{MLPF} algorithm improves the physics response with respect to a rule-based benchmark algorithm and demonstrates computationally scalable \acl{PF} reconstruction in a high-\acl{PU} environment.
\end{abstract}


\acresetall 

\section{Introduction}
Reconstruction algorithms at general-purpose high-energy particle detectors aim to provide a holistic, well-calibrated physics interpretation of the collision event. 
Variants of the \ac{PF} algorithm have been used at the CELLO~\cite{Behrend:1982gk}, ALEPH~\cite{Buskulic:1994wz}, H1~\cite{H1:2020zpd}, ZEUS~\cite{Breitweg:1997aa,Breitweg:1998gc}, DELPHI~\cite{Abreu:1995uz}, CDF~\cite{Bocci:2001zx,Connolly:2003gb,Abulencia:2007iy}, D0~\cite{Abazov:2008ff}, CMS~\cite{Sirunyan:2017ulk} and ATLAS~\cite{Aaboud:2017aca} experiments to reconstruct a particle-level interpretation of high-multiplicity hadron collision events, given individual detector elements such as tracks and calorimeter clusters from a multi-layered, heterogeneous, irregular-geometry detector. 
The \ac{PF} algorithm generally correlates tracks and calorimeter clusters from detector layers such as the \ac{ECAL}, \ac{HCAL} and others to reconstruct charged and neutral hadron candidates as well as photons, electrons, and muons with an optimized efficiency and resolution. 
Existing \ac{PF} reconstruction implementations are tuned using simulation for each specific experiment because detailed detector characteristics and geometry are critical for the best possible physics performance.

Recently, there has been significant interest in adapting the \ac{PF} reconstruction approach for future high-luminosity experimental conditions at the \ac{LHC}~\cite{Chlebana:2203028}, as well as for proposed future collider experiments such as the \ac{FCC}~\cite{Selvaggi:2715344,Benedikt:2018csr}. 
\Ac{PF} reconstruction is also a key driver in the detector design for future lepton colliders~\cite{Abada:2019zxq,Behnke:2013xla,CEPCStudyGroup:2018ghi}.
While reconstruction algorithms are often based on an imperative, rule-based approach, the use of supervised \ac{ML} to define reconstruction parametrically based on data and simulation samples may improve the physics reach of the experiments by allowing a more detailed reconstruction to be deployed given a fixed computing budget.
Reconstruction algorithms based on \ac{ML} may be well-suited to irregular, high-granularity detector geometries and for novel signal models, where it may not be feasible to encode the necessary granularity in the ruleset.
A fully probabilistic particle-level interpretation of the event from an \ac{ML}-based reconstruction may also improve the physics performance of downstream algorithms such as jet tagging with more granular inputs.
At the same time, \ac{ML}-solutions for computationally intensive problems may offer a modern computing solution that may scale better with the expected progress on \ac{ML}-specific computing infrastructures, e.g., at high-performance computing centers.

\ac{ML}-based reconstruction approaches using \acp{GNN}~\cite{gnn,gilmer2017neural,pointnet,Battaglia:2016jem,DGCNN,pointnet} have been proposed for various tasks in particle physics~\cite{Shlomi:2020gdn}, including tracking~\cite{Farrell:2018cjr,Ju:2020xty,Amrouche:2019wmx,Amrouche:2019yxv,Choma:2020cry}, jet finding~\cite{Ju:2020tbo,Li:2020grn,Guo:2020vvt} and tagging~\cite{Moreno:2019bmu,Moreno:2019neq,Qu:2019gqs,Mikuni:2020wpr}, calorimeter reconstruction~\cite{Qasim:2019otl}, pileup mitigation~\cite{Martinez:2018fwc}, and \ac{PF} reconstruction~\cite{Kieseler:2020wcq,DiBello:2020bas,Duarte:2020ngm}. 
The clustering of energy deposits in detectors with a realistic, irregular-geometry detector using \acp{GNN} has been first proposed in Ref.~\cite{Qasim:2019otl}. 
The \ac{ML}-based reconstruction of overlapping signals without a regular grid was further developed in Ref.~\cite{Kieseler:2020wcq}, where an optimization scheme for reconstructing a variable number of particles based on a potential function using an object condensation approach was proposed. 
The clustering of energy deposits from particle decays with potential overlaps is an essential input to \ac{PF} reconstruction. 
In Ref.~\cite{DiBello:2020bas}, various \ac{ML} models including \acp{GNN} and computer-vision models have been studied for reconstructing neutral hadrons from multi-layered granular calorimeter images and tracking information. 
In particle gun samples, the \ac{ML}-based approaches achieved a significant improvement in neutral hadron energy resolution over the default algorithm, which is an important step towards a fully parametric, simulation-driven reconstruction using \ac{ML}.

In this paper, we build on the previous \ac{ML}-based reconstruction approaches by extending the \ac{ML}-based \ac{PF} algorithm to reconstruct particle candidates in events with a large number of simultaneous \ac{PU} collisions.
In \cref{sec:sim}, we propose a benchmark dataset that has the main components for a particle-level reconstruction of charged and neutral hadrons with \ac{PU}. 
In \cref{sec:model}, we propose a \ac{GNN}-based \ac{MLPF} algorithm where the runtime scales approximately linearly with the input size. 
Furthermore, in \cref{sec:results}, we characterize the performance of the \ac{MLPF} model on the benchmark dataset in terms of hadron reconstruction efficiency, fake rate and resolution, comparing it to the baseline \ac{PF} reconstruction, while also demonstrating using synthetic data that \ac{MLPF} reconstruction can be computationally efficient and scalable. 
Finally, in \cref{sec:outlook} we discuss some potential issues and next steps for \ac{ML}-based \ac{PF} reconstruction.

\section{Physics simulation}
\label{sec:sim}
We use \PYTHIA8~\cite{Sjostrand:2006za,Sjostrand:2007gs} and \DELPHES3~\cite{deFavereau:2013fsa} from the HepSim software repository~\cite{Chekanov:2014fga} to generate a particle-level dataset of 50,000 top quark-antiquark (\ttbar) events produced in proton-proton collisions at 14\TeV, overlaid with minimum bias events corresponding to a \ac{PU} of 200 on average. 
The \ttbar dataset is used for training the \ac{MLPF} model. 
We additionally generate 5,000 events composed uniquely of jets produced through the strong interaction, referred to as quantum chromodynamics (QCD) multijet events, with the same \ac{PU} conditions for validation to evaluate the model in a different physics regime from the training dataset.
The dataset consists of detector hits as the input, generator particles as the ground truth and reconstructed particles from \DELPHES for additional validation.
The QCD sample uses a minimum invariant $\pt$ of 20\GeV, otherwise, the same generator settings are used as for the \ttbar sample. 
The \DELPHES model corresponds to a CMS-like detector with a multi-layered charged particle tracker, an electromagnetic and hadron calorimeter.
The full \PYTHIA8 and \DELPHES data cards are available on Zenodo along with the dataset~\cite{pata_joosep_2021_4452283}.

Although this simplified simulation does not include important physics effects such as pair production, Brehmsstrahlung, nuclear interactions, electromagnetic showering or a detailed detector simulation, it allows the study of overall per-particle reconstruction properties for charged and neutral hadrons in a high-\ac{PU} environment. 
Different reconstruction approaches can be developed and compared on this simplified dataset, where the expected performance is straightforward to assess, including from the aspect of computational complexity.

The inputs to \ac{PF} are charged particle tracks and calorimeter clusters. 
We use these high-level detector inputs (elements), rather than low-level tracker hits or unclustered calorimeter hits to closely follow how \ac{PF} is implemented in existing reconstruction chains, where successive reconstruction steps are decoupled, such that each step can be optimized and characterized individually. 
In this toy dataset, tracks are characterized by transverse momentum ($\pt$)~\footnote{As common for collider physics, we use a Cartesian coordinate system with the $z$ axis oriented along the beam axis, the $x$ axis on the horizontal plane, and the $y$ axis oriented upward. 
The $x$ and $y$ axes define the transverse plane, while the $z$ axis identifies the longitudinal direction. 
The azimuthal angle $\phi$ is computed with respect to the $x$ axis. 
The polar angle $\theta$ is used to compute the pseudorapidity $\eta = -\log(\tan(\theta/2))$. 
The transverse momentum ($\pt$) is the projection of the particle momentum on the ($x$, $y$) plane. 
We fix units such that $c=\hslash=1$.}, charge, and the pseudorapidity and azimuthal angle coordinates ($\eta, \phi$), including extrapolations to the tracker edge ($\eta_\mathrm{outer}, \phi_\mathrm{outer}$).

The track $\eta$ and $\phi$ coordinates are additionally smeared with a 1\% Gaussian resolution to model a finite tracker resolution.
Calorimeter clusters are characterized by electromagnetic or hadron energy $E$ and $\eta,\phi$ coordinates. 
In this simulation, an event has $N=(4.9 \pm 0.3) \times 10^{3}$ detector inputs on average.

\begin{figure*}[t!]
\centering
\includegraphics[width=\textwidth]{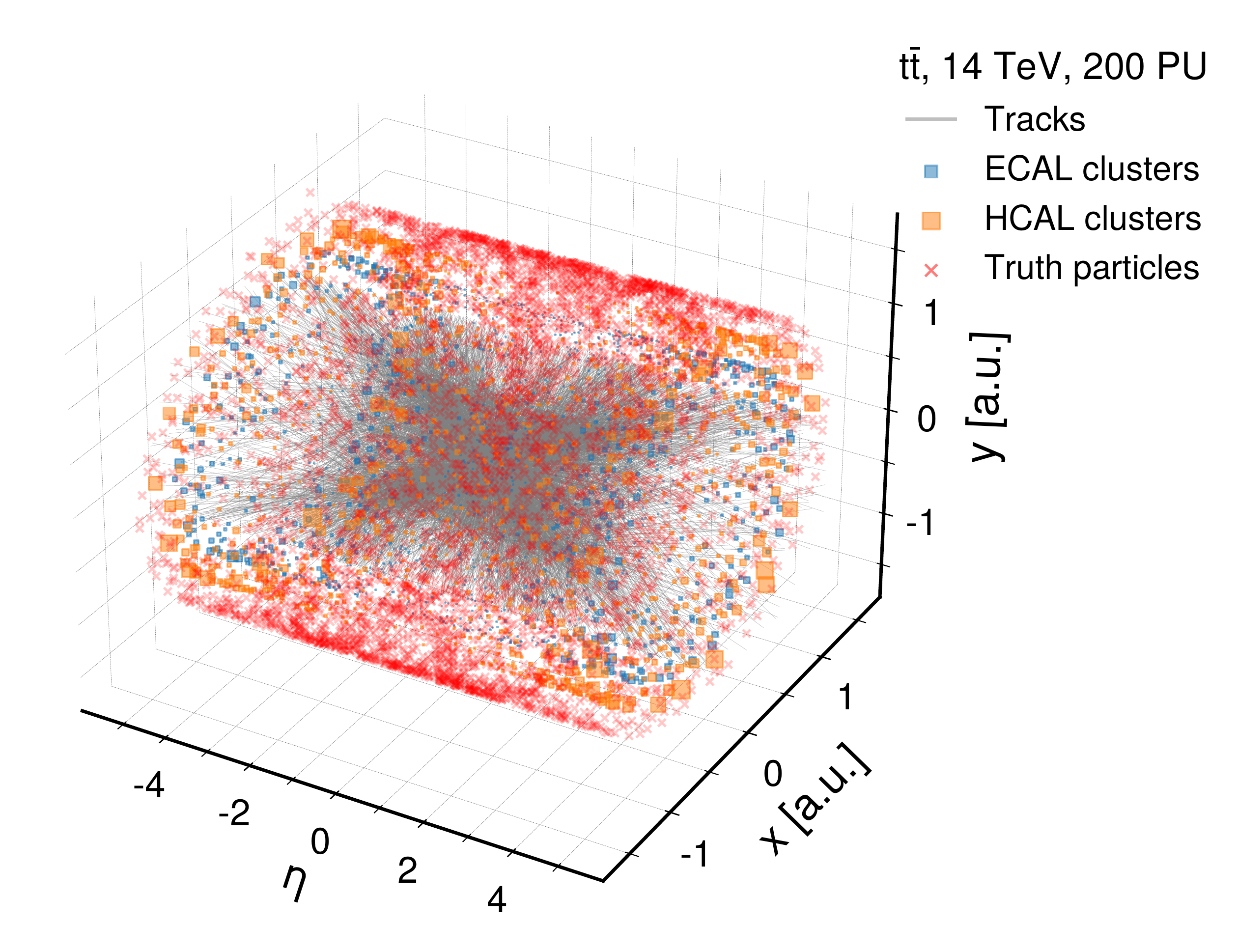}
\caption{A simulated \ttbar event from the \ac{MLPF} dataset with 200 \ac{PU} interactions. 
The input tracks are shown in gray, with the trajectory curvature being defined by the inner and outer $\eta, \phi$ coordinates. 
Electromagnetic (hadron) calorimeter clusters are shown in blue (orange), with the size corresponding to cluster energy for visualization purposes. 
We also show the locations of the generator particles (all types) with red cross markers. 
The radii and thus the $x,y$-coordinates of the tracker, \ac{ECAL} and \ac{HCAL} surfaces are arbitrary for visualization purposes.}
\label{fig:event}
\end{figure*}

\begin{figure}[t!]
\centering
\includegraphics[width=\columnwidth]{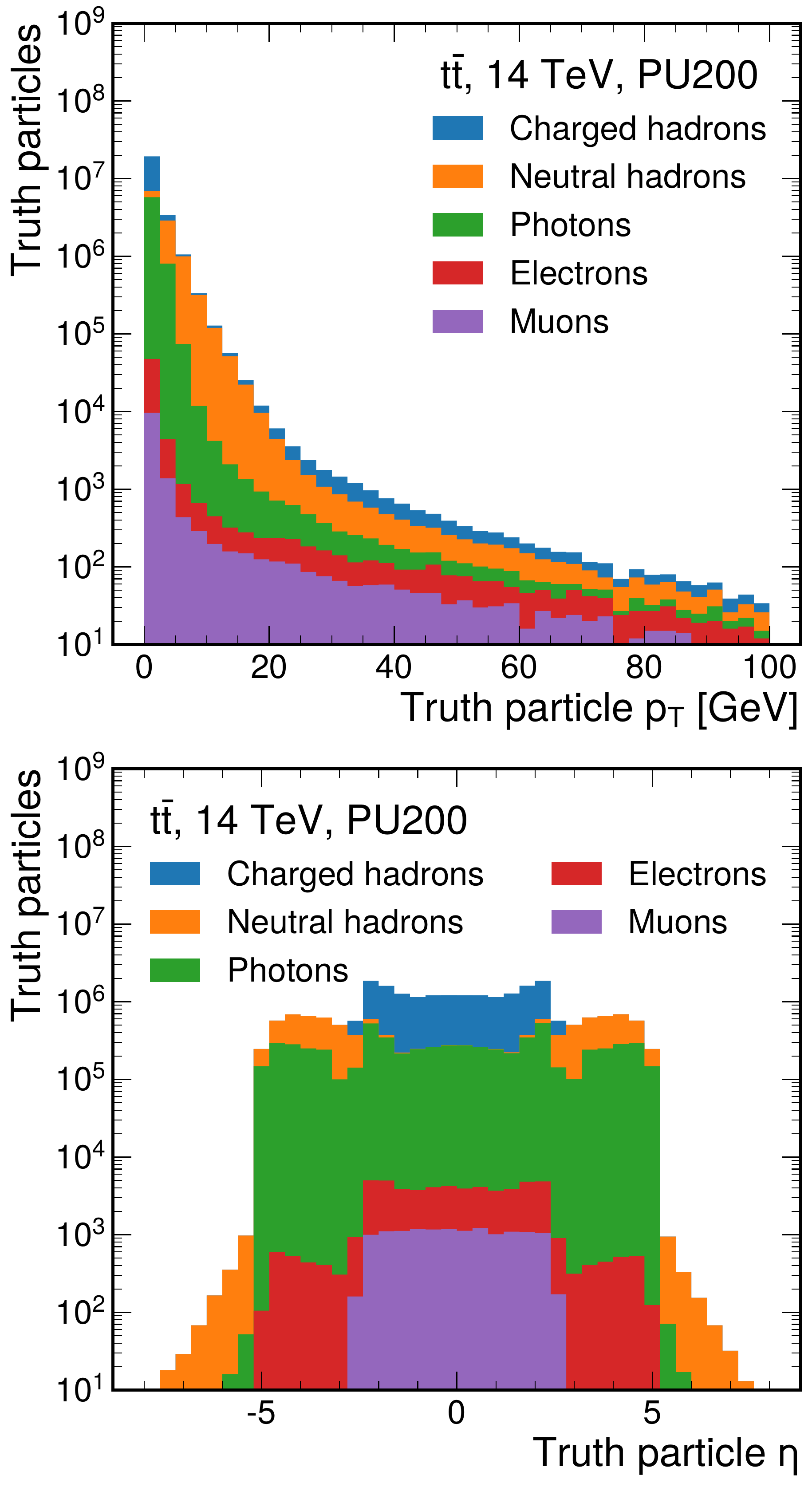}
\caption{The $\pt$ (upper) and $\eta$ (lower) distributions of the generator particles in the simulated \ttbar dataset with \ac{PU}, split by particle type.}
\label{fig:gen_pt_eta}
\end{figure}

The targets for \ac{PF} reconstruction are stable generator-level particles that are associated to at least one detector element, as particles that leave no detector hits are generally not reconstructable. 
Generator particles are characterized by a \ac{PID} which may take one of the following categorical values: charged hadron, neutral hadron, photon, electron, or muon. 
In case multiple generator particles all deposit their energy completely to a single calorimeter cluster, we treat them as reconstructable only in aggregate. 
In this case, the generator particles are merged by adding the momenta and assigning it the \ac{PID} of the highest-energy sub-particle. 
In addition, charged hadrons are indistinguishable outside the tracker acceptance from neutral hadrons, therefore we label generated charged hadrons with $|\eta| > 2.5$ to neutral hadrons. 
We also set a lower energy threshold on reconstructable neutral hadrons to $E > 9.0\GeV$ based on the \DELPHES rule-based \ac{PF} reconstruction, ignoring neutral hadrons that do not pass this threshold. 
A single event from the dataset is visualized in \cref{fig:event}, demonstrating the input multiplicity and particle distribution in the event. 
The differential distributions of the generator-level particles in the simulated dataset are shown in \cref{fig:gen_pt_eta}.

We also store the \ac{PF} candidates reconstructed by \DELPHES for comparison purposes. 
The \DELPHES rule-based PF algorithm is described in detail in Ref.~\cite{deFavereau:2013fsa}.
Charged and neutral hadrons are identified based on track and hadron calorimeter cluster overlaps and energy subtraction.
Photons are identified based on electromagnetic calorimeter clusters not matched to tracks.
In addition, we note that electrons and muons are identified by \DELPHES based on the generator particle associated to the corresponding track, therefore, for electron and muon tracks we add the corresponding generator-level identification as an input feature to the \ac{MLPF} training to demonstrate that given the appropriate detector inputs, these less common particles can also be identified by the algorithm.

Each event is now fully characterized by the set of generator particles $Y=\{y_j\}$ (target vectors), the set of detector inputs $X=\{x_i\}$ (input vectors), with
\begin{align}
y_j &= [\mathrm{PID}, \pt, E, \eta, \phi, q]\,,\\
x_i &= [\mathrm{type}, \pt, E_\mathrm{ECAL}, E_\mathrm{HCAL}, \eta, \phi, \eta_{\mathrm{outer}}, \phi_{\mathrm{outer}}, q]\,,\\
\mathrm{PID} &\in \{\mathrm{charged\ hadron}, \mathrm{neutral\ hadron}, \mathrm{\gamma}, \mathrm{e}^{\pm}, \mathrm{\mu}^{\pm}\}\,\\
\mathrm{type} &\in \{\mathrm{track}, \mathrm{cluster}\}\,.
\end{align}
For input tracks, only the type, $\pt$, $\eta$, $\phi$, $\eta_\mathrm{outer}$, $\phi_\mathrm{outer}$, and $q$ features are filled.
Similarly, for input clusters, only the type, $E_\mathrm{ECAL}$, $E_\mathrm{HCAL}$, $\eta$ and $\phi$ entries are filled.
Unfilled features for both tracks and clusters are set to zero.
In future iterations of \ac{MLPF}, it may be beneficial to represent input elements of different types with separate data matrices to improve the computational efficiency of the model.
Precomputing additional features such as track trajectory intersection points with the calorimeters may further improve the performance of \ac{PF} reconstruction based on machine learning.

Functionally, the detector is modelled in simulation by a function $S(Y)=X$ that produces a set of detector signals from the generator-level inputs for an event. 
Reconstruction imperfectly approximates the inverse of that function $R\simeq S^{-1}(X) = Y$. 
In the following section, we approximate the reconstruction as set-to-set translation and implement a baseline \ac{MLPF} reconstruction using \acp{GNN}.

\section{ML-based PF reconstruction}
\label{sec:model}
For a given set of detector inputs $X$, we want to predict a set of particle candidates $Y'$ that closely approximates the target generator particle set $Y$. 
The target and predicted sets may have a different number of elements, depending on the quality of the prediction. 
For use in \ac{ML} using gradient descent, this requires a computationally efficient, differentiable set-to-set metric $||Y - Y'|| \in \mathbb{R}$ to be used as the loss function. 

We simplify the problem numerically by first zero-padding the target set $Y$ such that $|Y|=|X|$. 
This turns the problem of predicting a variable number of particles into a multi-classification prediction by adding an additional ``no particle'' to the classes already defined by the target \ac{PID} and is based on Ref.~\cite{Kieseler:2020wcq}.
Furthermore, for \ac{PF} reconstruction, the target generator particles are often geometrically and energetically close to well-identifiable detector inputs. 
In physics terms, a charged hadron is reconstructed based on a track, while a neutral hadron candidate can always be associated to at least one primary source cluster, with additional corrections taken from other nearby detector inputs. 
Therefore, we choose to preprocess the inputs such that for a given arbitrary ordering of the detector inputs $X=[\dots, x_i, \dots]$ (sets of vectors are represented as matrices with some arbitrary ordering for \ac{ML} training), the target set $Y$ is arranged such that if a target particle can be associated to a detector input, it is arranged to be in the same location in the sequence. 
This data preprocessing step speeds up model convergence, but does not introduce any additional assumptions to the model.
Since the target set now has a predefined size, we may compute the loss function which approximates reconstruction quality element-by-element:
\begin{align}
||Y - Y'|| &\equiv \sum_{j \in \mathrm{event}} L(y_j,y'_j)\,,\\
L(y_j,y'_j) &\equiv \mathrm{CLS}(c_j, c'_j) + \alpha \mathrm{REG}(p_j, p'_j)\,,
\end{align}
where the target values and predictions $y_j = [c_j; p_j]$ are decomposed such that the multi-classification is encapsulated in the scores and one-hot encoded classes $c_j$, while the momentum and charge regression values in $p_j$. 
We use CLS to denote the multi-classification loss, while REG denotes the regression loss for the momentum components weighted appropriately by a coefficient $\alpha$.
This combined per-particle loss function serves as a baseline optimization target for the \ac{ML} training.
Further physics improvements may be reached by extending the loss to take into account event-level quantities, either by using an energy flow distance as proposed in Ref.~\cite{Komiske:2018cqr,Komiske:2019fks,Romao:2020ojy}, or using a particle-based~\cite{Kansal:2020svm,Bellagente:2019uyp,Belayneh:2019vyx,Butter:2019cae} \ac{GAN}~\cite{goodfellow2014generative} to optimize the reconstruction network in tandem with an adversarial classifier that is trained to distinguish between the target and reconstructed events, given the detector inputs.

\subsection{Graph neural network implementation}
\begin{figure*}[t!]
\centering
\includegraphics[width=\textwidth]{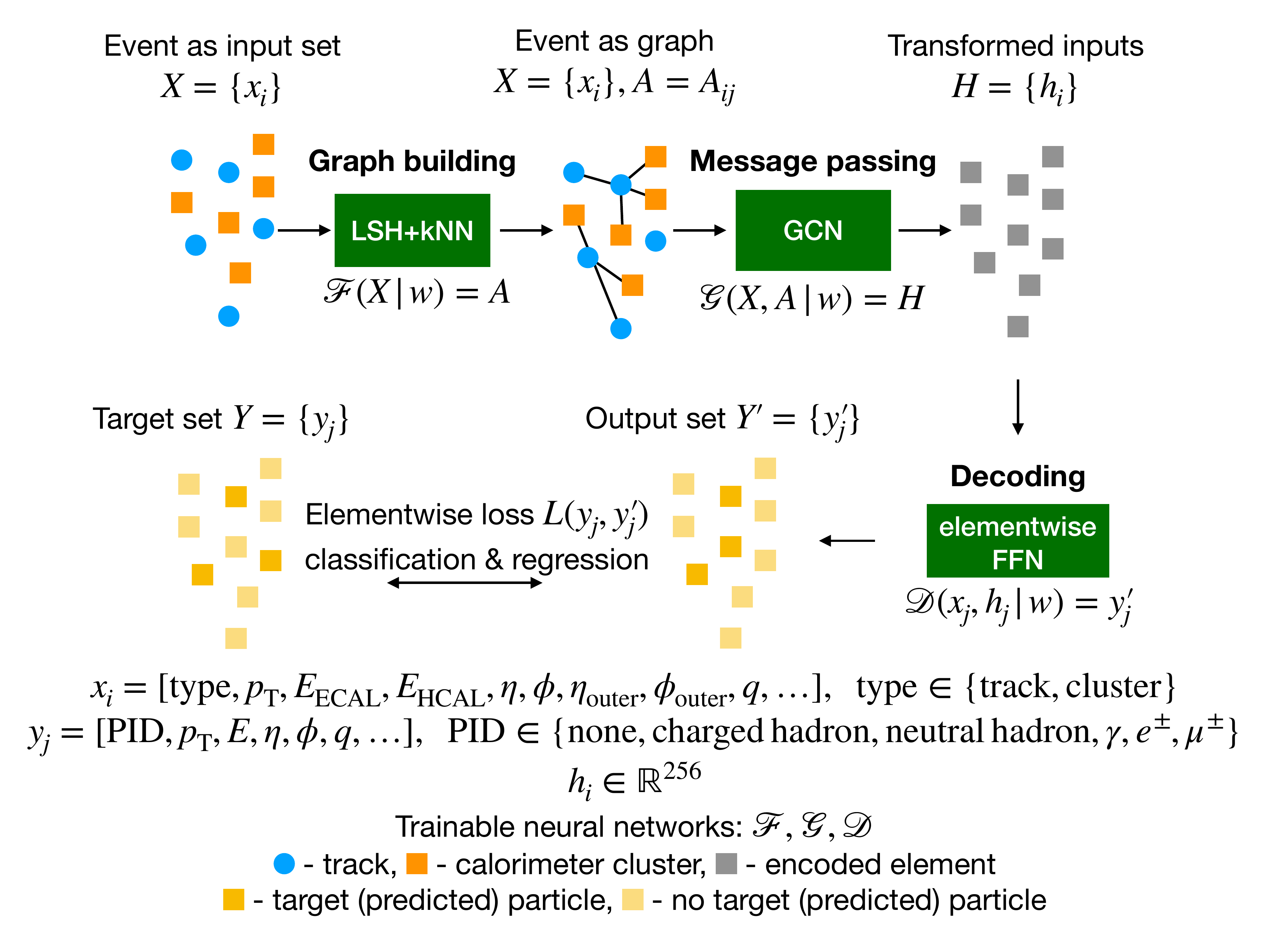}
\caption{Functional overview of the end-to-end trainable \ac{MLPF} setup with \acp{GNN}. 
The event is represented as a set of detector elements $x_i$. 
The set is transformed into a graph by the graph building step, which is implemented here using an \acf{LSH} approximation of \ac{kNN}. 
The graph nodes are then encoded using a message passing step, implemented using graph convolutional nets. 
The encoded elements are decoded to the output feature vectors $y_j$ using elementwise feedforward networks.}
\label{fig:schematic}
\end{figure*}

Given the set of detector inputs for the event $X=\{x_i\}$, we adopt a message passing approach for reconstructing the \ac{PF} candidates $Y=\{y_j\}$. 
First, we need to construct a trainable graph adjacency matrix $\mathcal{F}(X | w) = A$ for the given set of input elements, represented with the graph building block in \cref{fig:schematic}. 
The input set is heterogeneous, containing elements of different type (tracks, \ac{ECAL} clusters, \ac{HCAL} clusters) in different feature spaces. 
Therefore, defining a static neighborhood graph in the feature space in advance is not straightforward. 
A generic approach to learnable graph construction using \ac{kNN} in an embedding space, known as GravNet, has been proposed in Ref.~\cite{Qasim:2019otl}, where the authors demonstrated that a learnable, dynamically-generated graph structure significantly improves the physics performance of an \ac{ML}-based reconstruction algorithm for calorimeter clustering.
Similar dynamic graph approaches have also been proposed in Ref.~\cite{DGCNN}.

However, naive \ac{kNN} graph implementations in common \ac{ML} packages such as \TENSORFLOW or \pytorchgeometric have $\mathcal{O}(n^2)$ time complexity: for each set element out of $n=|X|$, we must order the other $n-1$ elements by distance and pick the $k$ closest. 
More efficient \ac{kNN} graph construction is possible with, for example, k-dimensional trees~\cite{rajani2015parallel}, but so far, we are not aware of an implementation that interfaces with common, differentiable \ac{ML} tools.
For reconstruction, given equivalent physics performance, both computational efficiency (a low overall runtime) and scalability (subquadratic time and memory scaling with the input size) are desirable.

We build on the GravNet approach~\cite{Qasim:2019otl} by using an approximate \ac{kNN} graph construction algorithm based on \acf{LSH} to improve the time complexity of the graph building algorithm. 
The \ac{LSH} approach has been recently proposed~\cite{kitaev2020reformer} for approximating and thus speeding up \ac{ML} models that take into account element-to-element relations using an optimizable $n \times n$ matrix known as self-attention~\cite{vaswani2017attention}.
The method divides the input into bins using a hash function, such that nearby elements are likely to be assigned to the same bin. 
The bins contain only a small number of elements, such that constructing a \ac{kNN} graph in the bin is significantly faster than for the full set of elements, and thus not strongly affected by the quadratic scaling of the \ac{kNN} algorithm.

In the \ac{kNN}+\ac{LSH} approach, the $n$ input elements $x_i$ are projected into a $d_K$-dimensional embedding space by a trainable, elementwise feed-forward network $\mathrm{FFN}(x_i | w) = z_i \in \mathbb{R}^{d_K}$. 
As in Ref.~\cite{kitaev2020reformer}, we now assign each element into one of $d_B$ bins indexed by integers $b_i$ using $h(z_i) = b_i \in [1, \dots, d_B]$, where $h(x)$ is a hash function that assigns nearby $x$ to the same bin with a high probability. 
We define the hash function as $h(x)=\argmax[xP; -xP]$ where $[u; v]$ denotes the concatenation of two vectors $u$ and $v$ and $P$ is a random projection matrix of size $[d_K, d_B/2]$ drawn from the normal distribution at initialization.

We now build $d_B$ \ac{kNN} graphs based on the embedded elements $z_i$ in each of the \ac{LSH} bins, such that the full sparse graph adjacency $A_{ij}$ in the inputs set $X$ is defined by the sum of the subgraphs. 
The embedding function can be optimized with backpropagation and gradient descent using the values of the nonzero elements of $A_{ij}$. 
Overall, this graph building approach has $\mathcal{O}(n \log{n})$ time complexity and does not require the allocation of an $n^2$ matrix at any point. 
The \ac{LSH} step generates $d_B$ disjoint subgraphs in the full event graph. 
This is motivated by physics, as we expect subregions of the detector to be reconstructable approximately independently. 
The existing \ac{PF} algorithm in the CMS detector employs a similar approach by producing disjoint \ac{PF} blocks as an intermediate step of the algorithm~\cite{Sirunyan:2017ulk}. 

Having built the graph dynamically, we now use a variant of message passing~\cite{4700287,Battaglia:2016jem,gilmer2017neural,battaglia2018relational} to create hidden encoded states $\mathcal{G}(x_i, A_{ij} | w) = h_i$ of the input elements taking into account the graph structure. 
As a first baseline, we use a variant of \ac{GCN} that combines local and global node-level information~\cite{kipf2016semi,wu2019simplifying,xin2020graph}. 
This choice is motivated by implementation and evaluation efficiency in establishing a baseline. 
This message passing step is represented in \cref{fig:schematic} by the \ac{GCN} block. 
Finally, we decode the encoded nodes $H=\{h_i\}$ to the target outputs with an elementwise feed-forward network that combines the hidden state with the original input element $\mathcal{D}(x_i, h_i | w) = y'_i$ using a skip connection.

We have a joint graph building, but separate graph convolution and decoding layers for the multi-classification and the momentum and charge regression subtasks. 
This allows each subtask to be retrained separately in addition to a combined end-to-end training should the need arise. 
The classification and regression losses are combined with constant empirical weights such that they have an approximately equal contribution to the full training loss.
We use categorical cross-entropy for the classification loss, which measures the similarity between the true label distribution $c_j$ and the predicted labels $c'_j$.
For the regression loss, we use componentwise mean-squared error between the true and predicted momenta, where the losses for the individual momentum components $(\pt, \eta, \sin{\phi}, \cos{\phi}, E)$ are scaled by normalization factors such that the components have approximately equal contributions to the total loss.
It may be beneficial to use specific multi-task training strategies such as gradient surgery~\cite{yu2020gradient} to further improve the performance across all subtasks and to reduce the reliance on ad-hoc scale factors between the losses in a multi-task setup.

The multi-classification prediction outputs for each node are converted to particle probabilities with the softmax operation. 
We choose the \ac{PID} with the highest probability for the reconstructed particle candidate, while ensuring that the probability meets a threshold that matches a fake rate working point defined by the baseline \DELPHES \ac{PF} reconstruction algorithm.

The predicted graph structure is an intermediate step in the model and is not used in the loss function explicitly---we only optimize the model with respect to reconstruction quality. 
However, using the graph structure in the loss function when a known ground truth is available may further improve the optimization process. 
In addition, access to the predicted graph structure may be helpful in evaluating the interpretability of the model.

The set of networks for graph building, message passing and decoding has been implemented with \TENSORFLOW2.3 and can be trained end-to-end using gradient descent. 
The inputs are zero-padded to $n=6,400$ elements.
Additional elements beyond 6,400 are truncated for efficient training and performance evaluation, amounting to about 0.007\% of the total number of elements in the \ttbar simulation sample. 
The truncated elements are always calorimeter towers as the order of the elements is set by the \DELPHES simulation. 
For inference during data taking, truncation should be avoided.
The \ac{LSH} bin size chosen to be $128$ such that the number of bins $d_B=50$ and the number of nearest neighbors $k=16$. 
We use two hidden layers for each encoding and decoding net with 256 units each, with two successive graph convolutions between the encoding and decoding steps. 
Exponential linear activations (ELU)~\cite{clevert2016fast} are used for the hidden layers and linear activations are used for the outputs. 
Overall, the model has approximately 1.5 million trainable weights and 25,000 constant weights for the random projections. 
For optimization, we use the Adam~\cite{kingma2017adam} algorithm with a learning rate of $5\times10^{-6}$ for 300 epochs, training over $4\times 10^4$ events, with $10^4$ events used for testing. 
The events are processed in minibatches of five simultaneous events per \ac{GPU}, we train for approximately 48 hours using five RTX 2070S \acp{GPU} using data parallelism on 40,000 simulated \ttbar events.
We report the results of the multi-task learning problem in the next section. 
The code and dataset to reproduce the training are made available on the Zenodo platform~\cite{joosep_pata_2021_4452542,pata_joosep_2021_4452283}.

\section{Results}
\label{sec:results}
\begin{figure*}[t!]
\centering
\includegraphics[width=\textwidth]{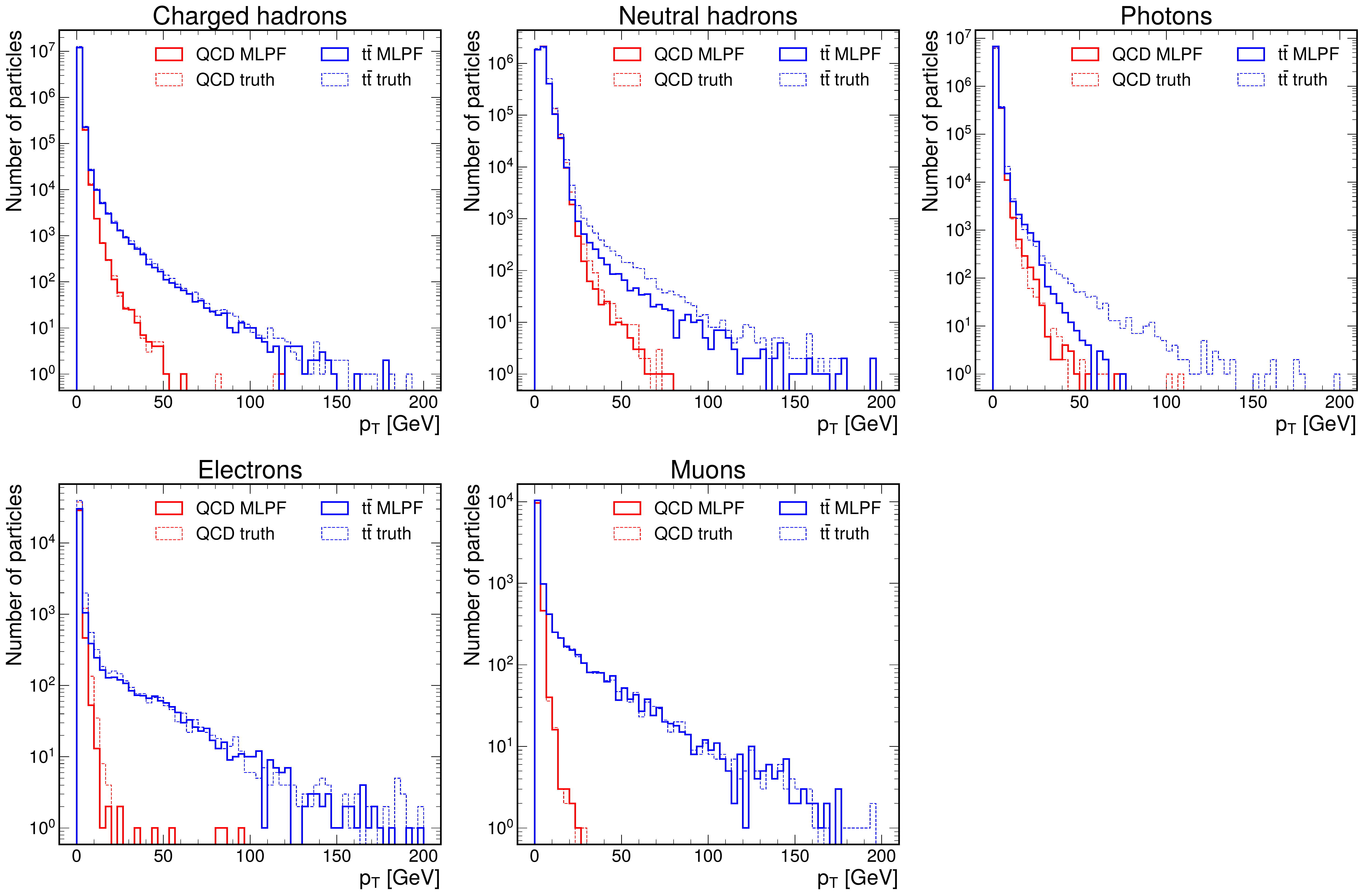}
\caption{The \ac{MLPF} reconstruction compared to the truth-level $\pt$ distribution for the QCD validation sample and the \ttbar sample used for training. 
The differences between the \ac{MLPF} and truth distributions are a measure of the prediction error. 
Charged hadrons, electrons, and muons are identified based on tracks with no misidentification or loss of efficiency, hence the prediction error is negligible for both samples. 
For neutral hadrons and photons, the tail is reconstructed at a lower efficiency for \ttbar as compared to QCD, which could arise from overrepresentation of low-$\pt$ particles in the unweighted \ttbar training sample.}
\label{fig:qcd_vs_ttbar}
\end{figure*}

In \cref{fig:qcd_vs_ttbar}, we show the $\pt$ distributions for the \ac{MLPF} reconstruction and generator-level truth for both simulated QCD multijet and \ttbar events.
Although the \ac{MLPF} model was trained on \ttbar, we observe a slight underprediction at high transverse momentum for photons and neutral hadrons, which could arise from the much greater numbers of low-$\pt$ particles relative to high-$\pt$ particles in this unweighted sample.
Further work is needed to improve the performance in the high-$\pt$ tail of the distribution.
We find that the model generalizes well to the QCD sample that was not used in the training, demonstrating that the \ac{MLPF}-based reconstruction is transferable across different physics samples.

\begin{figure}[t!]
\centering
\includegraphics[width=\columnwidth]{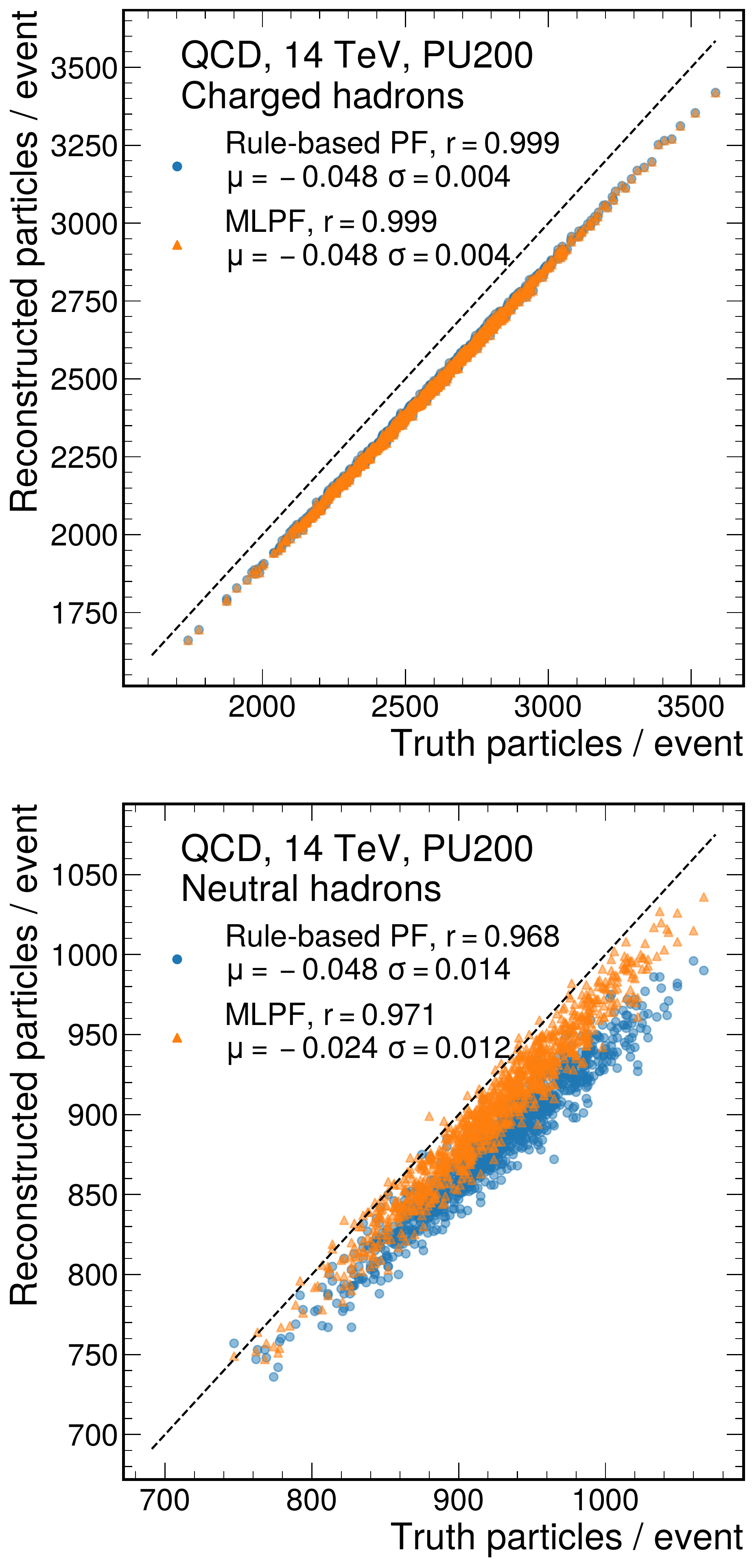}
\caption{True and predicted particle multiplicity for \ac{MLPF} and \DELPHES \ac{PF} for charged (upper) and neutral hadrons (lower) in simulated QCD multijet events with \ac{PU}. 
Both models show a high degree of correlation ($r$) between the generated and predicted particle multiplicity, with the \ac{MLPF} model reconstructing the neutral particle multiplicities with improved resolution ($\sigma$) and a lower bias ($\mu$).}
\label{fig:num}
\end{figure}

For the following results, we focus on the charged and neutral hadron performance in QCD events, as hadrons make up the bulk of the energy content of the jets and thus are the primary target for \ac{PF} reconstruction.
We do not report detailed performance characteristics for photons, electrons, and muons at this time because of the limitations of the \DELPHES dataset and the rule-based \ac{PF} algorithm.
A realistic study of photon and electron disambiguation, in particular, requires a more detailed dataset that includes additional physics effects, as discussed in \cref{sec:sim}.
In \cref{fig:num}, we present the charged and neutral hadron multiplicities from both the baseline rule-based \ac{PF} and \ac{MLPF} algorithms as a function of the target multiplicities. 
The particle multiplicities from the \ac{MLPF} model correlate better with the generator-level target than the rule-based \ac{PF} algorithm, demonstrating that the multi-classification model successfully reconstructs variable-multiplicity events.
In general, we do not observe significant differences in the physics performance of the \ac{MLPF} algorithm between the QCD and \ttbar samples in the phase space where we have validated it.

\begin{figure}[t!]
\centering
\includegraphics[width=\columnwidth]{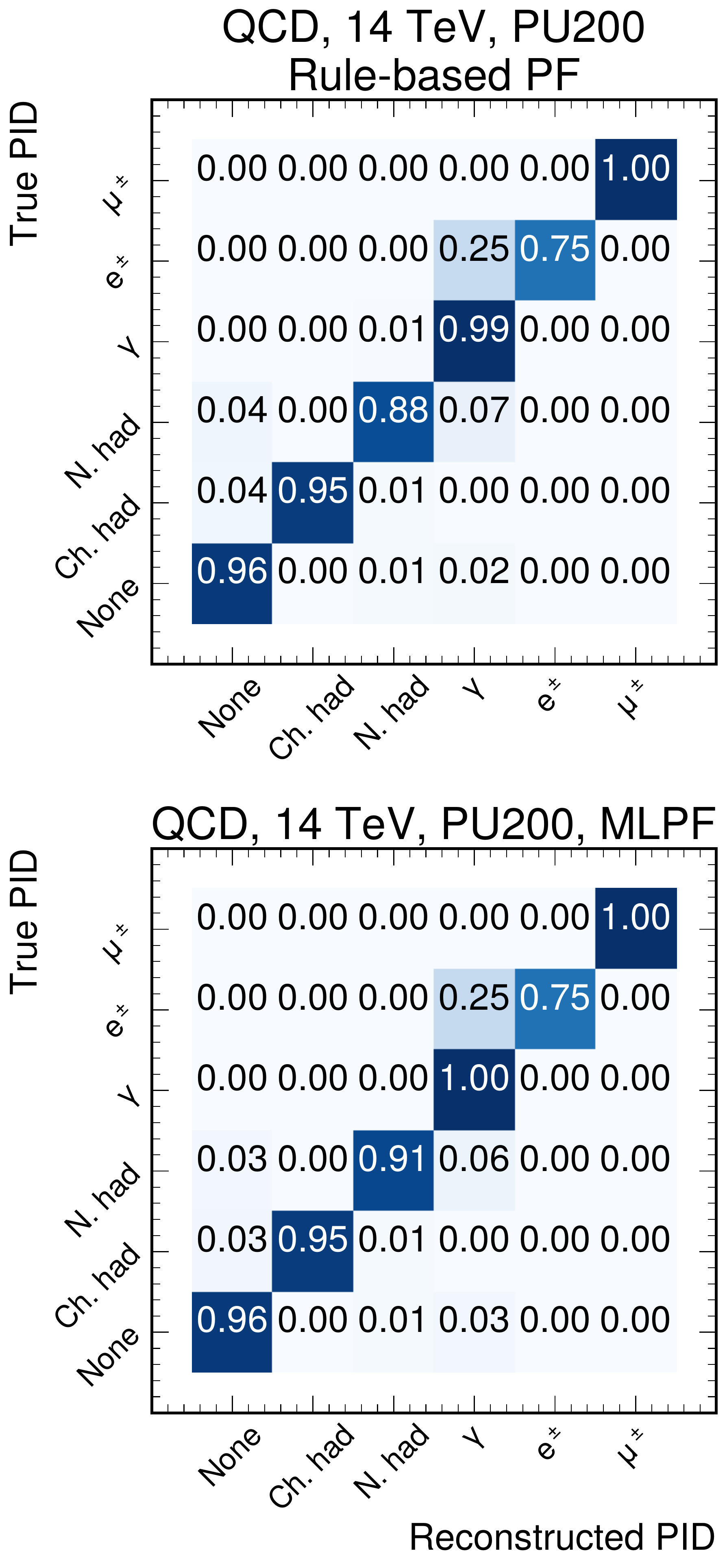}
\caption{Particle identification confusion matrices in simulated QCD multijet events with \ac{PU}, with gen-level particles as the ground truth, showing the baseline rule-based \DELPHES \ac{PF} (upper) and the \ac{MLPF} (lower) outputs.
The rows have been normalized to unit probability, corresponding to normalizing the dataset according to the generated \ac{PID}.}
\label{fig:confusion}
\end{figure}

In \cref{fig:confusion}, we compare the per-particle multi-classification confusion matrix for both reconstruction methods. 
We see overall a similar classification performance for both approaches. The charged hadron identification performance is driven by track efficiency and is the same for \ac{MLPF} and the rule-based \ac{PF}. The neutral hadron identification efficiency is slightly higher for \ac{MLPF} (0.91 vs 0.88), since hadron calorimeter cluster energies that are not matched to tracks must be determined algorithmically for neutral hadron reconstruction.
The electron-photon misidentification is driven by the parametrized tracking efficiency, as electromagnetic calorimeter clusters without an associated track are reconstructed as photons.
Electron and muon identification performance is shown simply for completeness, as it is driven by the use of generator-level \ac{PID} values for those tracks.
Improved Monte Carlo generation, subsampling, or weighting may further improve reconstruction performance for particles or kinematic configurations that occur rarely in a physical simulation. 
In this set of results, we apply no weighting on the events or particles in the event.

\begin{figure}[t!]
\centering
\includegraphics[width=\columnwidth]{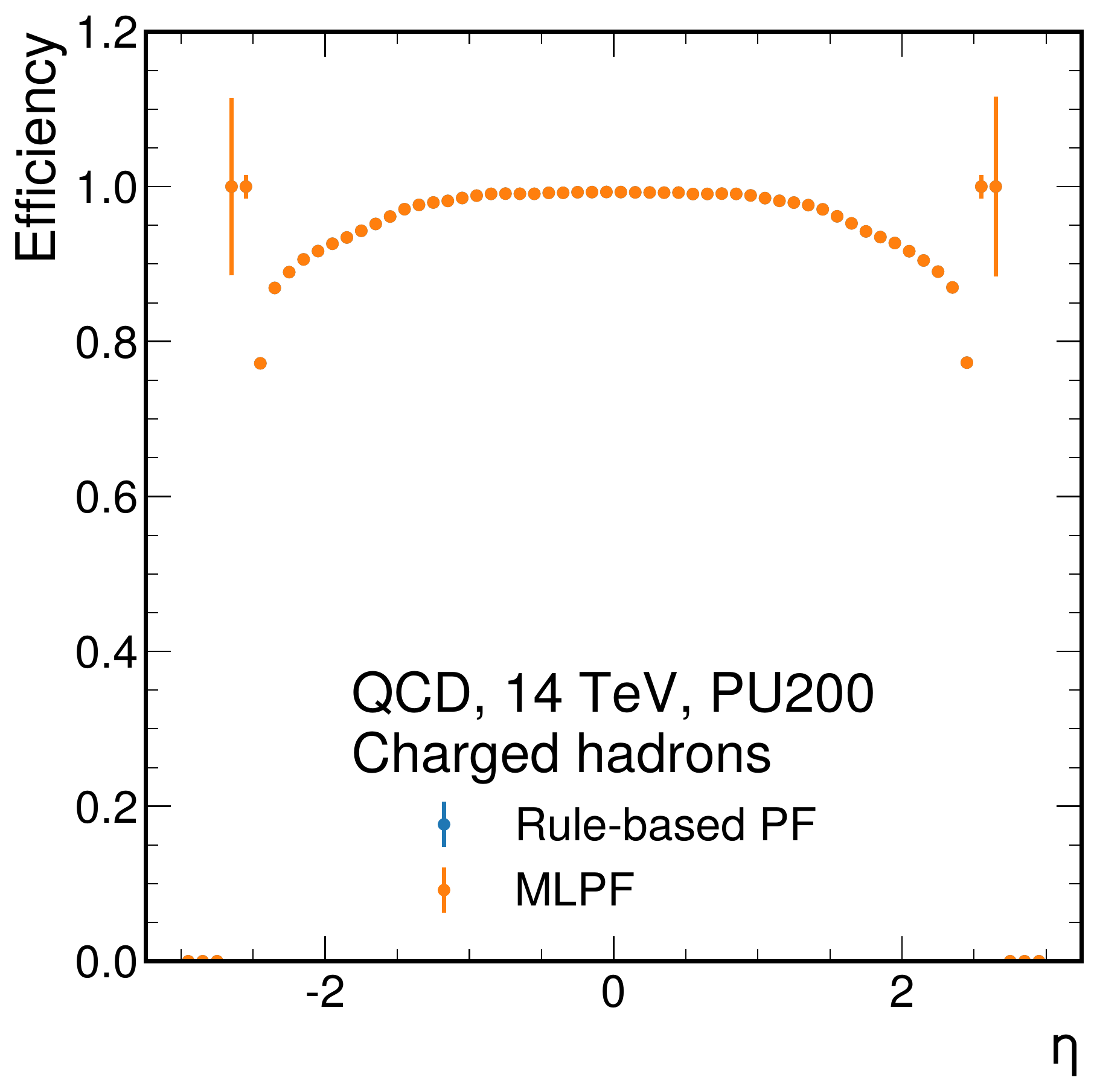}
\caption{The efficiency of reconstructing charged hadron candidates as a function of the generator particle pseudorapidity $\eta$ in simulated QCD multijet events with \ac{PU}. 
Since the simulation does not contain fake tracks, the charged hadron reconstruction is driven entirely by tracking efficiency and is the same for \ac{MLPF} and the rule-based \ac{PF}.}
\label{fig:eff_fake_pid1}
\end{figure}

\begin{figure}[t!]
\centering
\includegraphics[width=\columnwidth]{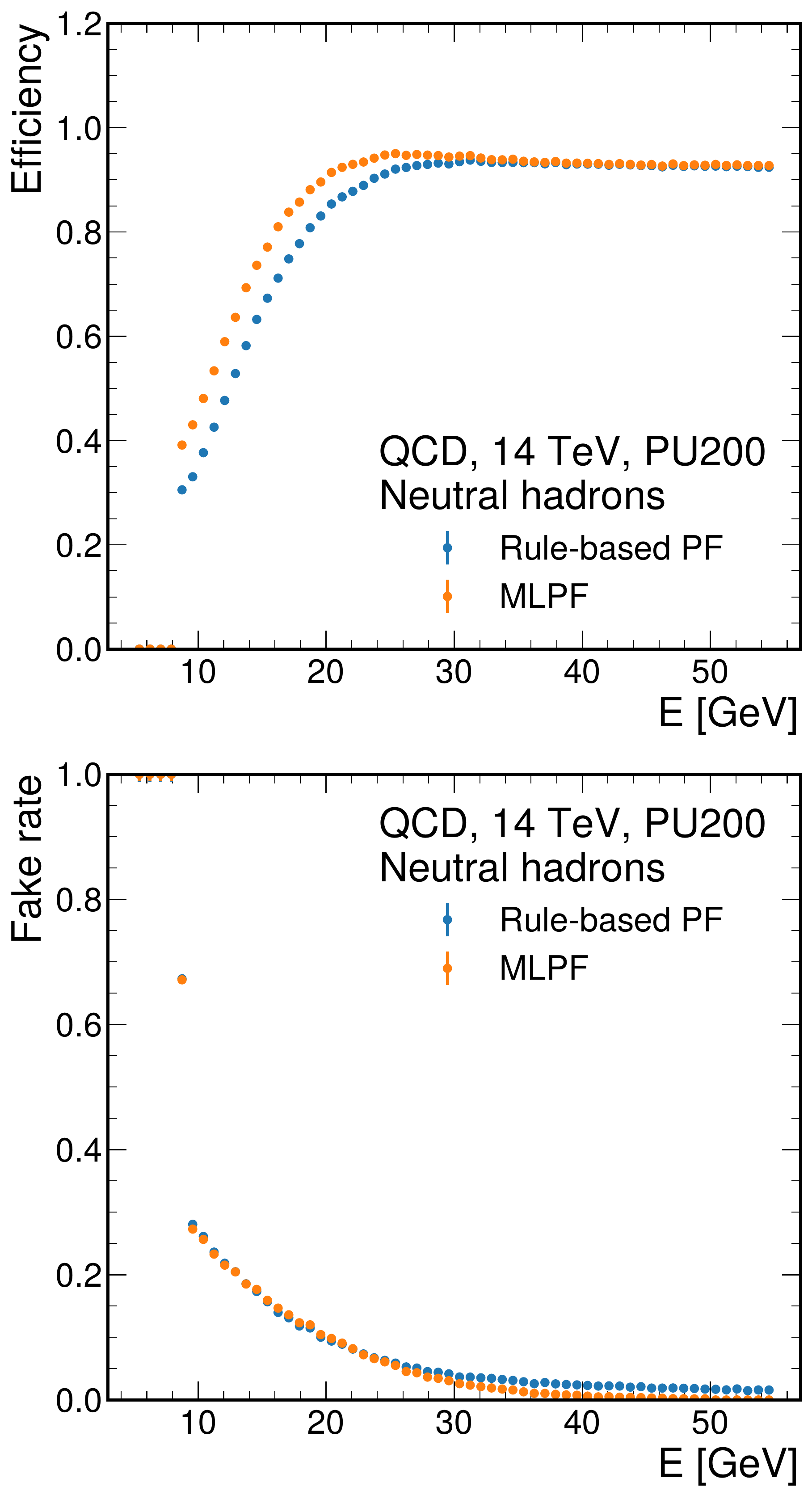}
\caption{The efficiency (upper) and fake rate (lower) of reconstructing neutral hadron candidates as a function of the generator particle energy in simulated QCD multijet events with \ac{PU}. 
The \ac{MLPF} model shows comparable performance to the \DELPHES \ac{PF} benchmark, with a somewhat lower fake rate at a similar efficiency.}
\label{fig:eff_fake_pid2}
\end{figure}

In \cref{fig:eff_fake_pid1}, we see that the $\eta$-dependent charged hadron efficiency (true positive rate) for the \ac{MLPF} model is somewhat higher than for the rule-based \ac{PF} baseline, while the fake rate (false positive rate) is equivalently zero, as the \DELPHES simulation includes no fake tracks. 
From \cref{fig:eff_fake_pid2}, we observe a similar result for the energy-dependent efficiency and fake rate of neutral hadrons. 
Both algorithms exhibit a turn-on at low energies and show a constant behaviour at high energies, with \ac{MLPF} being comparable or slightly better than the rule-based \ac{PF} baseline.

\begin{figure}[t!]
\centering
\includegraphics[width=\columnwidth]{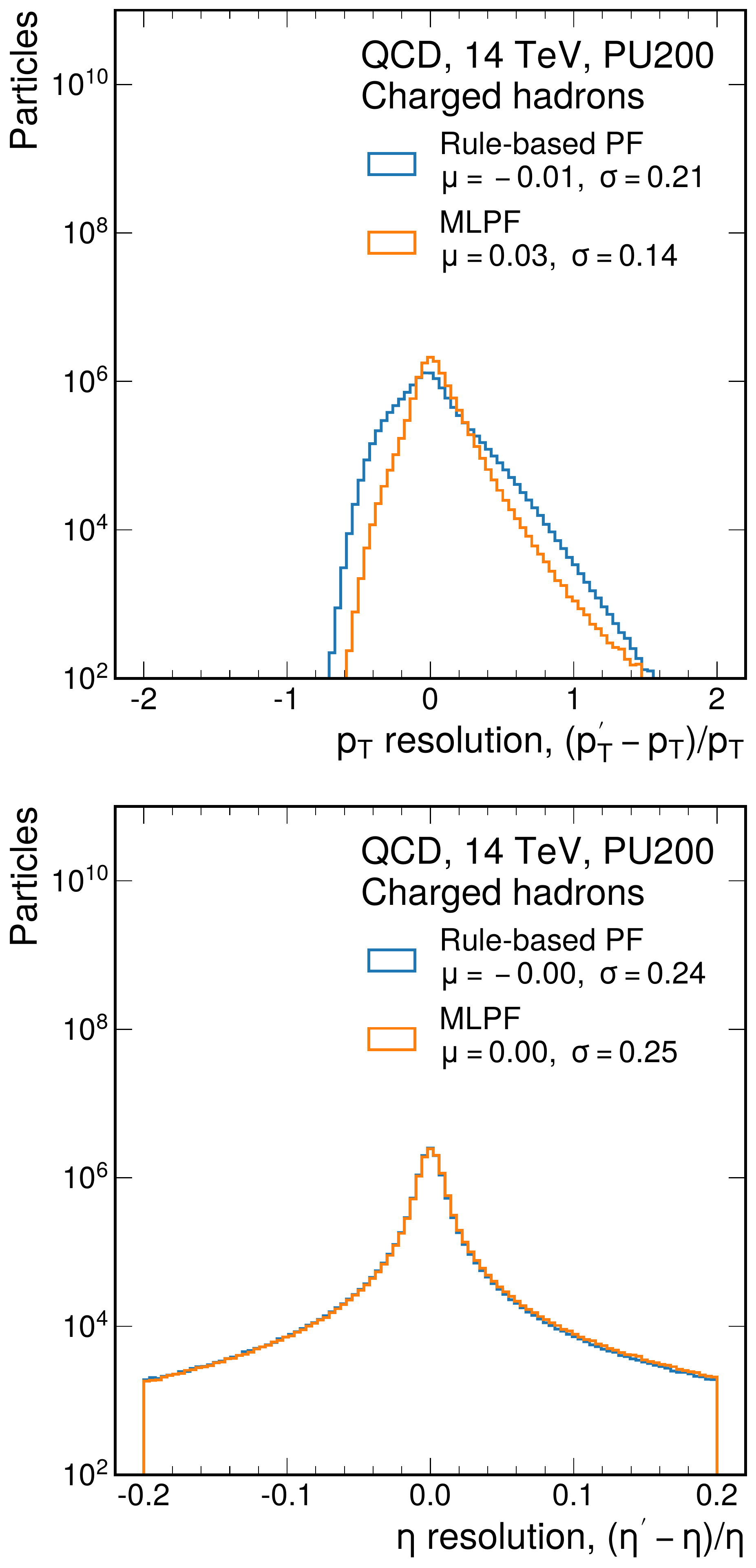}
\caption{The $\pt$ and $\eta$ resolution of the \DELPHES \ac{PF} benchmark and the \ac{MLPF} model for charged hadrons in simulated QCD multijet events with \ac{PU}. 
The $\pt$ resolution is comparable for both algorithms, with the angular resolution being driven by the smearing of the track $(\eta, \phi)$ coordinates.}
\label{fig:res_pid1}
\end{figure}

\begin{figure}[t!]
\centering
\includegraphics[width=\columnwidth]{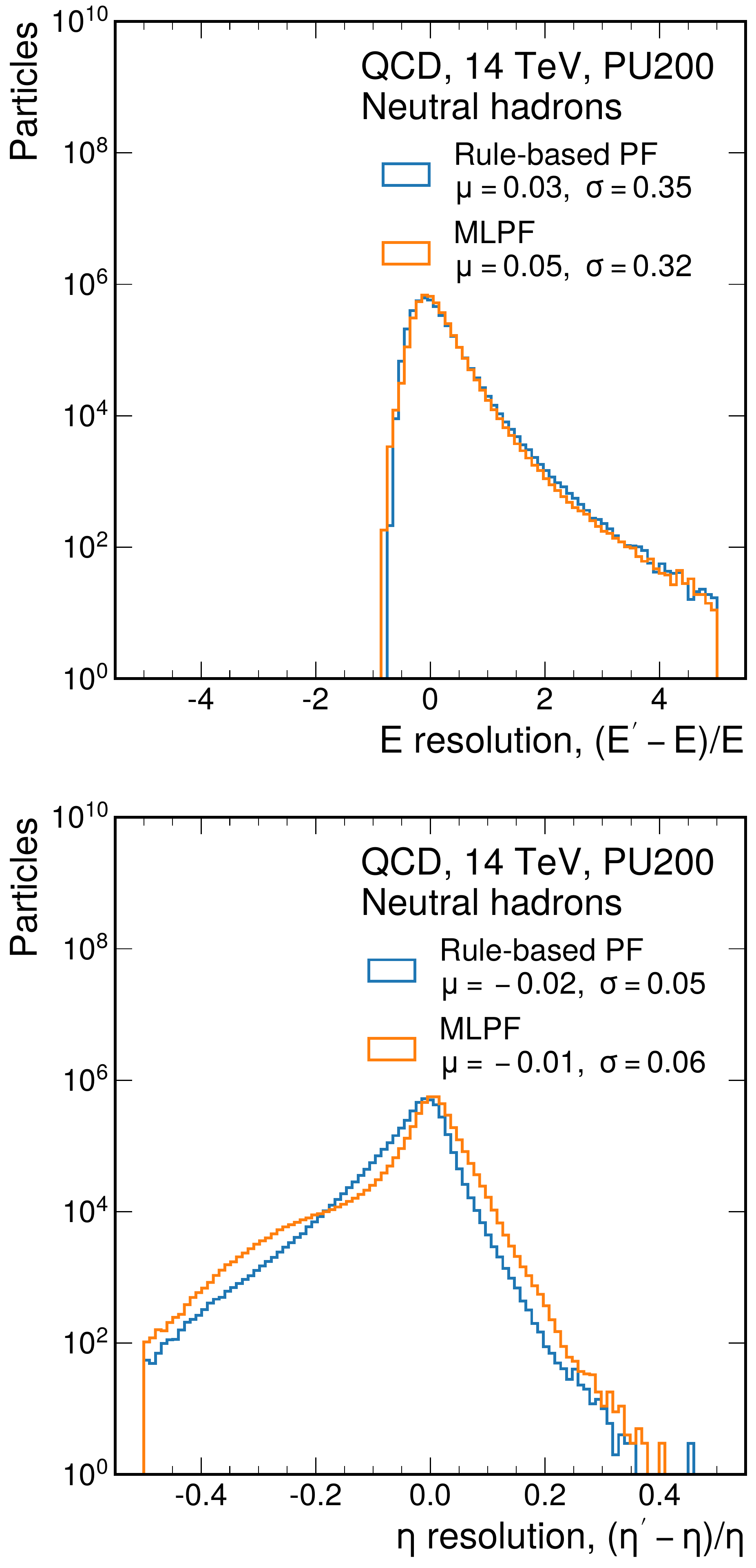}
\caption{The energy and $\eta$ resolution of the \DELPHES \ac{PF} benchmark and the \ac{MLPF} model for neutral hadrons in simulated QCD multijet events with \ac{PU}. 
Both reconstruction algorithms show comparable performance.}
\label{fig:res_pid2}
\end{figure}

Furthermore, we see on \cref{fig:res_pid1,fig:res_pid2} that the energy, energy (\pt) and angular resolution of the \ac{MLPF} algorithm are generally comparable to the baseline for neutral (charged) hadrons. 

Overall, these results demonstrate that formulating \ac{PF} reconstruction as a multi-task \ac{ML} problem of simultaneously identifying charged and neutral hadrons in a high-\ac{PU} environment and predicting their momentum may offer comparable or improved physics performance over hand-written algorithms in the presence of sufficient simulation samples and careful optimization.
The performance characteristics for the baseline and the proposed \ac{MLPF} model are summarized in~\cref{tab:results}.

\begin{figure}[t!]
\centering
\includegraphics[width=\columnwidth]{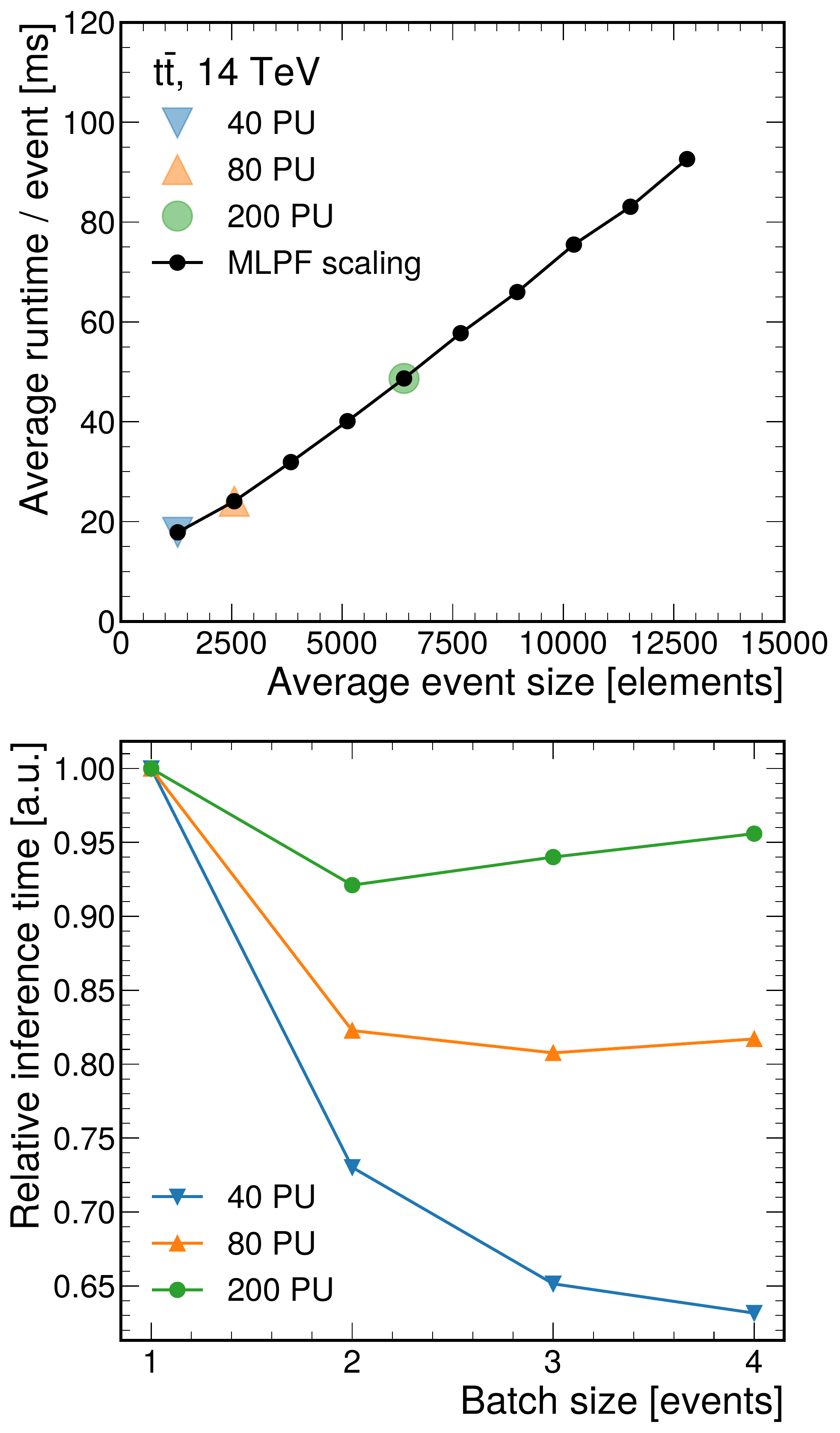}
\caption{Average runtime of the \ac{MLPF} \ac{GNN} model with a varying input event size (upper) and the relative inference time when varying the number of events evaluated simultaneously, i.e. batch size (lower), normalized to batch size 1.
For a simulated event equivalent to 200 \ac{PU} collisions, we see a runtime of around 50\unit{ms}, which scales approximately linearly with respect to the input event size.
We see a weak dependence on batch size, with batching having a minor positive effect for low-\acl{PU} events. 
The runtime for each event size is averaged over 100 randomly generated events over three independent runs. 
The timing tests were done using an Nvidia RTX 2060S \ac{GPU} and an Intel i7-10700@2.9GHz CPU.
We assume a linear scaling between \ac{PU} and the number of detector elements.}
\label{fig:inference_timing}
\end{figure}

We also characterize the computational performance of the \ac{GNN}-based \ac{MLPF} algorithm. 
In \cref{fig:inference_timing}, we see that the average inference time scales roughly linearly with the input size, which is necessary for scalable reconstruction at high \ac{PU}. 
We also note that the \ac{GNN}-based \ac{MLPF} algorithm runs natively on a \ac{GPU}, with the current runtime at around 50\unit{ms/event} on a consumer-grade \ac{GPU} for a full 200 \ac{PU} event. 
The algorithm is simple to port to computing architectures that support common \ac{ML} frameworks like \TENSORFLOW without significant investment.
This includes \acp{GPU} and potentially even \acp{FPGA} or \ac{ML}-specific processors such as the GraphCore \acp{IPU}~\cite{Mohan:2020vvi} through specialized \ac{ML} compilers~\cite{Duarte:2018ite,Iiyama:2020wap,Heintz:2020soy}.
These coprocessing accelerators can be integrated into existing CPU-based experimental software frameworks as a scalable service that grows to meet the transient demand~\cite{Duarte:2019fta,Krupa:2020bwg,Rankin:2020usv}.

\begin{table}[t!]
    \centering
    \resizebox{\columnwidth}{!}{
    \begin{tabular}{c|cc|cc}
        & \multicolumn{2}{c}{Charged hadrons} & \multicolumn{2}{c}{Neutral hadrons} \\
        Metric & Rule-based PF & \ac{MLPF} & Rule-based PF & \ac{MLPF} \\
        \hline
        Efficiency & 0.953 & 0.953 & 0.883 & \textbf{0.908} \\
        Fake rate & 0.000 & 0.000 & 0.071 & \textbf{0.068} \\
        $\pt$ ($E$) resolution & 0.213 & \textbf{0.137} & 0.350 & \textbf{0.323} \\
        $\eta$ resolution & \textbf{0.240} & 0.245 & \textbf{0.050} & 0.058 \\
        $N$ resolution & 0.004 & 0.004 & 0.014 & \textbf{0.013} \\
    \end{tabular}}
    \caption{Particle reconstruction efficiency and fake rate, multiplicity $N$, $\pt$ ($E$) and $\eta$ resolutions for charged (neutral) hadrons, comparing the rule-based \ac{PF} baseline and the proposed \ac{MLPF} method.
    Bolded values indicate better performance.}
    \label{tab:results}
\end{table}

\section{Discussion and outlook}
\label{sec:outlook}
\acresetall

We have developed a \ac{ML} algorithm for \ac{PF} reconstruction in a high-\acl{PU} environment for a general-purpose multilayered particle detector based on transforming input sets of detector elements to the output set of reconstructed particles. 
The \ac{MLPF} implementation with \acp{GNN} is based on graph building with a \ac{LSH} approximation for \ac{kNN}, dubbed \ac{LSH}+\ac{kNN}, and message passing using graph convolutions. 
Based on benchmark particle-level \ttbar and QCD multijet datasets generated using \PYTHIA8 and \DELPHES3, the \ac{MLPF} \ac{GNN} reconstruction offers comparable performance to the baseline rule-based \ac{PF} algorithm in \DELPHES, demonstrating that a purely parametric \ac{ML}-based \ac{PF} reconstruction can reach or exceed the physics performance of existing reconstruction algorithms, while allowing for greater portability across various computing architectures at a possibly reduced cost. 
The inference time empirically scales approximately linearly with the input size, which is useful for efficient evaluation in the high-luminosity phase of the \ac{LHC}. 
In addition, the \ac{ML}-based reconstruction model may offer useful features for downstream physics analysis like per-particle probabilities for different reconstruction interpretations, uncertainty estimates, and optimizable particle-level reconstruction for rare processes including displaced signatures.

The \ac{MLPF} model can be further improved with a more physics-motivated optimization criterion, i.e. a loss function that takes into account event-level, in addition to particle-level differences. 
While we have shown that a per-particle loss function already converges to an adequate physics performance overall, improved event-based losses such as the object condensation approach or energy flow may be useful. 
In addition, an event-based loss may be defined using an adversarial classifier that is trained to distinguish the target particles from the reconstructed particles. 

Reconstruction algorithms need to adapt to changing experimental conditions---this may be addressed in \ac{MLPF} by a periodic retraining on simulation that includes up-to-date running condition data such as the beam-spot location, dead channels, and latest calibrations.
In a realistic \ac{MLPF} training, care must be taken that the reconstruction qualities of rare particles and particles in the low-probability tails of distributions are not adversely affected and that the reconstruction performance remains uniform. 
This may be addressed with detailed simulations and weighting schemes. 
In addition, for a reliable physics result, the interpretability of the reconstruction is essential.
The reconstructed graph structure can provide information about causal relations between the input detector elements and the reconstructed particle candidates.

In order to develop a usable \ac{ML}-based \ac{PF} reconstruction algorithm, a realistic high-\acl{PU} simulated dataset that includes detailed interactions with the detector material needs to be used for the \ac{ML} model optimization.
The model should be optimized and validated on a mix of realistic high-\ac{PU} events to learn global properties of reconstruction, as well as on a set of particle gun samples to ensure that local properties of particle reconstruction are learned in a generalizable way.
To evaluate the reconstruction performance, efficiencies, fake rates, and resolutions for all particle types need to be studied in detail as a function of particle kinematics and detector conditions. 
Furthermore, high-level derived quantities such as \acl{PU}-dependent jet and missing transverse momentum resolutions must be assessed for a more complete characterization of the reconstruction performance. 
With ongoing work in \ac{ML}-based track and calorimeter cluster reconstruction upstream of \ac{PF}~\cite{CERN-LHCC-2017-023,ATL-PHYS-PUB-2020-018,deOliveira:2018lqd,Belayneh:2019vyx,Ju:2020xty,Choma:2020cry} and \ac{ML}-based reconstruction of high-level objects including jets and jet classification probabilities downstream of \ac{PF}~\cite{Sirunyan:2017ezt,Moreno:2019bmu,Moreno:2019neq,Qu:2019gqs,Aad:2019uoz,Aad:2019aic,Bols:2020bkb,Sirunyan:2020lcu}, care must be taken that the various steps are optimized and interfaced coherently.

Finally, the \ac{MLPF} algorithm is inherently parallelizable and can take advantage of hardware acceleration of \acp{GNN} via \acp{GPU}, \acp{FPGA} or emerging \ac{ML}-specific processors.
Current experimental software frameworks can easily integrate coprocessing accelerators as a scalable service.
By harnessing heterogeneous computing and parallelizable, efficient \ac{ML}, the burgeoning computing demand for event reconstruction tasks in the high-luminosity \ac{LHC} era can be met while maintaining or even surpassing the current physics performance.

\begin{acknowledgements}
We would like to thank Guenther Dissertori for suggesting the idea of \ac{ML}-driven \ac{PF} reconstruction several years ago in private discussions.
We thank our colleagues in the CMS Collaboration, especially in the Particle Flow, Physics Performance and Datasets, Offline and Computing, and Machine Learning groups, in particular Josh Bendavid, Kenichi Hatakeyama, Lindsey Gray, Jan Kieseler, Danilo Piparo, Gregor Kasieczka, Laurits Tani, and Juska Pekkanen, for helpful feedback in the course of this work.

J.~P. was supported by the Prime National Science Foundation (NSF) Tier2 award 1624356 and the U.S. Department of Energy (DOE), Office of Science, Office of High Energy Physics under Award No. DE-SC0011925 while at Caltech, and is currently supported by the Mobilitas Pluss Grant No. MOBTP187 of the Estonian Research Council.
J.~D. is supported by the DOE, Office of Science, Office of High Energy Physics Early Career Research program under Award No. DE-SC0021187 and by the DOE, Office of Advanced Scientific Computing Research under Award No. DE-SC0021396 (FAIR4HEP).
M.~P. is supported by the European Research Council (ERC) under the European Union's Horizon 2020 research and innovation program (Grant Agreement No. 772369).
J-R.~V. and M.~S. are supported by the DOE, Office of Science, Office of High Energy Physics under Award No. DE-SC0011925, DE-SC0019227, and DE-AC02-07CH11359.
J-R.~V. was additionally partially supported the same ERC grant as M.~P.

We are grateful to Caltech and the Kavli Foundation for their support of undergraduate student research in cross-cutting areas of machine learning and domain sciences.
This work was conducted at ``\textit{iBanks},'' the AI GPU cluster at Caltech, and on the NICPB GPU resources, supported by European Regional Development Fund through the CoE program grant TK133.
We acknowledge Nvidia, SuperMicro and the Kavli Foundation for their support of \textit{iBanks}.
Part of this work was also performed using the Pacific Research Platform Nautilus HyperCluster supported by NSF awards CNS-1730158, ACI-1540112, ACI-1541349, OAC-1826967, the University of California Office of the President, and the University of California San Diego's California Institute for Telecommunications and Information Technology/Qualcomm Institute. 
Thanks to CENIC for the 100\unit{Gpbs} networks.
\end{acknowledgements}

\bibliographystyle{lucas_unsrt_epjc}
\bibliography{main}
\end{document}